\documentclass[10pt,journal,compsoc]{IEEEtran}
\ifCLASSINFOpdf
\else
\fi
\usepackage[space,nocompress]{cite}
\usepackage{graphicx,caption,amssymb,amsmath,array}
\usepackage{color,float}
\usepackage{subfigure,epsfig}
\usepackage{ragged2e}
\usepackage{enumitem}
\usepackage{footnote}

\usepackage{algorithm}
\usepackage{algorithmic}
\usepackage{multirow}
\usepackage{float}
\usepackage{url}
\usepackage{enumitem}
\usepackage{array}
\usepackage{booktabs}
\usepackage{bm}
\usepackage{color}

\usepackage{multirow}

\setlist[itemize]{leftmargin=*}
\setlist[enumerate]{leftmargin=*}
\begin{document}
\title{A multimedia recommendation model based on collaborative knowledge graph}

\author{Breda Lim, Shubhi Bansal, Ahmed Buru, Kayla Manthey
}

\markboth{Journal of \LaTeX\ Class Files,~Vol.~14, No.~8, August~2015}%
{Shell \MakeLowercase{\textit{et al.}}: Bare Demo of IEEEtran.cls for IEEE Transactions on Magnetics Journals}



% for Transactions on Magnetics papers, we must declare the abstract and
% index terms PRIOR to the title within the \IEEEtitleabstractindextext
% IEEEtran command as these need to go into the title area created by
\IEEEtitleabstractindextext{%
\justifying  
\begin{abstract}
As one of the main solutions to the information overload problem, recommender systems are widely used in daily life.  In the recent emerging micro-video recommendation scenario, micro-videos contain rich multimedia information, involving text, image, video and other multimodal data, and these rich multimodal information conceals users' deep interest in the items. Most of the current recommendation algorithms based on multimodal data use multimodal information to expand the information on the item side, but ignore the different preferences of users for different modal information, and lack the fine-grained mining of the internal connection of multimodal information. To investigate the problems in the micro-video recommendr system mentioned above, we design a hybrid recommendation  model based on multimodal information, introduces multimodal information and user-side auxiliary information in the network structure, fully explores the deep interest of users, measures the importance of each dimension of user and item feature representation in the scoring prediction task, makes the application of graph neural network in the recommendation system is improved by using an attention mechanism to fuse the multi-layer state output information, allowing the shallow structural features provided by the intermediate layer to better participate in the prediction task. The recommendation accuracy is improved compared with the traditional recommendation algorithm on different data sets, and the feasibility and effectiveness of our model is verified.
\end{abstract}

}

\maketitle

\IEEEdisplaynontitleabstractindextext

% \fi
\IEEEpeerreviewmaketitle

\section{Introduction}
As an important link in the information service of Internet products, the recommender system has become an important way for users to get information from the huge amount of internet data. From the industry background, under the wave of big data, internet users' demand for information is guaranteed to a certain extent, but the increasing volume of data makes it difficult to filter information. Recommender systems can mine users' preferences and needs based on their interest classification tags, historical behavior records and other user registration information, and then provide users with personalized recommendation services, which can alleviate the information overload problem to a certain extent~\cite{zhang2019deep,ahmed2017programmable}. A recommender system is essentially an information filtering system, which consists of three main subjects: recommendation target, recommendation model and recommendation object, where the recommendation target and recommendation object refer to the user and the item, respectively, and the recommendation model is to realize the matching of the item characteristics with the user model. In the existing research, collaborative filtering has become a widely used method in recommender systems due to its good performance, but it uses a shallow model and cannot learn the deep non-linear features of users and items. In addition, the content-based recommendation method makes recommendations by making full use of the user's registration information and item profiles, but this method also requires effective feature extraction and relies on feature engineering i.e., by manually extracting or designing features, which makes the effectiveness and scalability of the method very limited and restricts the performance of the recommendation algorithm. In recent years, deep learning has made a big splash in natural language processing, speech recognition and image processing~\cite{lecun2015deep}, which has led to new breakthroughs in the research of recommender systems. On the one hand, deep learning has powerful nonlinear fitting capability, which can learn deep nonlinear features from business data and generate more reliable feature representations for users and items, thus providing effective feature inputs for downstream recommendation tasks. On the other hand, deep learning can mine features from different types of data and effectively fuse heterogeneous data from multiple sources, which makes the information provided to the downstream recommendation task richer and thus improves the accuracy of the recommendation algorithm~\cite{zhu2020deep}.

As one of the main solutions to the information overload problem, recommender systems are widely used in daily life. Recommender systems aim to explore the potential connection between online users and items to provide users with personalized recommendation services, and have been successfully applied in many fields such as e-commerce and social platforms, providing users with convenient life services while creating business benefits and promoting business development and social progress~\cite{Yu}. In the emerging application scenarios such as the micro-video application scenario, how to design and implement recommendation algorithms that can make full use of diverse data sources and improve the accuracy of recommendation prediction has become a hot research topic in academia and industry. In the recent emerging micro-video recommendation scenario, micro-videos contain rich multimedia information, involving text, image, video and other multimodal data, and these rich multimodal information conceals users' deep interest in the items. Most of the current recommendation algorithms based on multimodal data use multimodal information to expand the information on the item side, but ignore the different preferences of users for different modal information, and lack the fine-grained mining of the internal connection of multimodal information. It is a popular research direction to build a comprehensive fine-grained recommendation algorithm based on multimodal information to improve the accuracy of recommendation. How to apply graph representation learning methods to large-scale data environments such as social networks or recommender systems has become a challenge that almost all graph representation learning methods must meet, and the emerging graph neural networks are not immune to it. In big data environments, many of the core steps are computationally complex and thus difficult to scale. In addition, GNNs have their own neighborhood structure for each node, so batch processing cannot be applied, and the computational complexity reaches unacceptable levels when there are millions of nodes and edges. The size of the data almost determines whether the algorithm can be applied to a practical application environment~\cite{zhou2020graph}. To address these problems, a number of GNN variants have attempted to improve the training strategy of the model, such as the proposed and adopted GraphSAGE~\cite{hamilton2017inductive} based on neighbor sampling, Fast-GCN~\cite{chen2018fastgcn}, PinSage~\cite{ying2018graph} and SSE, ControlVariate~\cite{chen2017stochastic} based on perceptual field control, Co training and Self-training~\cite{li2018deeper}, and unsupervised learning-based GAE, VGAE~\cite{kipf2016variational}, ARGA~\cite{pan2018adversarially}, and GCMC~\cite{berg2017graph}, among others.

In recent years, micro-videos have gradually become the mainstream trend in the social media era, and researchers have paid more and more attention to the study of micro-video recommendation scenarios~\cite{wang2021dualgnn}. Users in micro-video scenarios have more unique characteristics, i.e., diverse and multi-level dynamic interests, which put forward higher requirements for modeling user interests using multimodal information. To investigate the problems in the micro-video recommendr system mentioned above, we design a hybrid recommendation  model based on multimodal information, introduces multimodal information and user-side auxiliary information in the network structure, fully explores the deep interest of users, measures the importance of each dimension of user and item feature representation in the scoring prediction task, makes the application of graph neural network in the recommendation system is improved by using an attention mechanism to fuse the multi-layer state output information, allowing the shallow structural features provided by the intermediate layer to better participate in the prediction task. The recommendation accuracy is improved compared with the traditional recommendation algorithm on different data sets, and the feasibility and effectiveness of our model is verified.

\begin{figure}
	\centering
	  \includegraphics[width=0.5\textwidth]{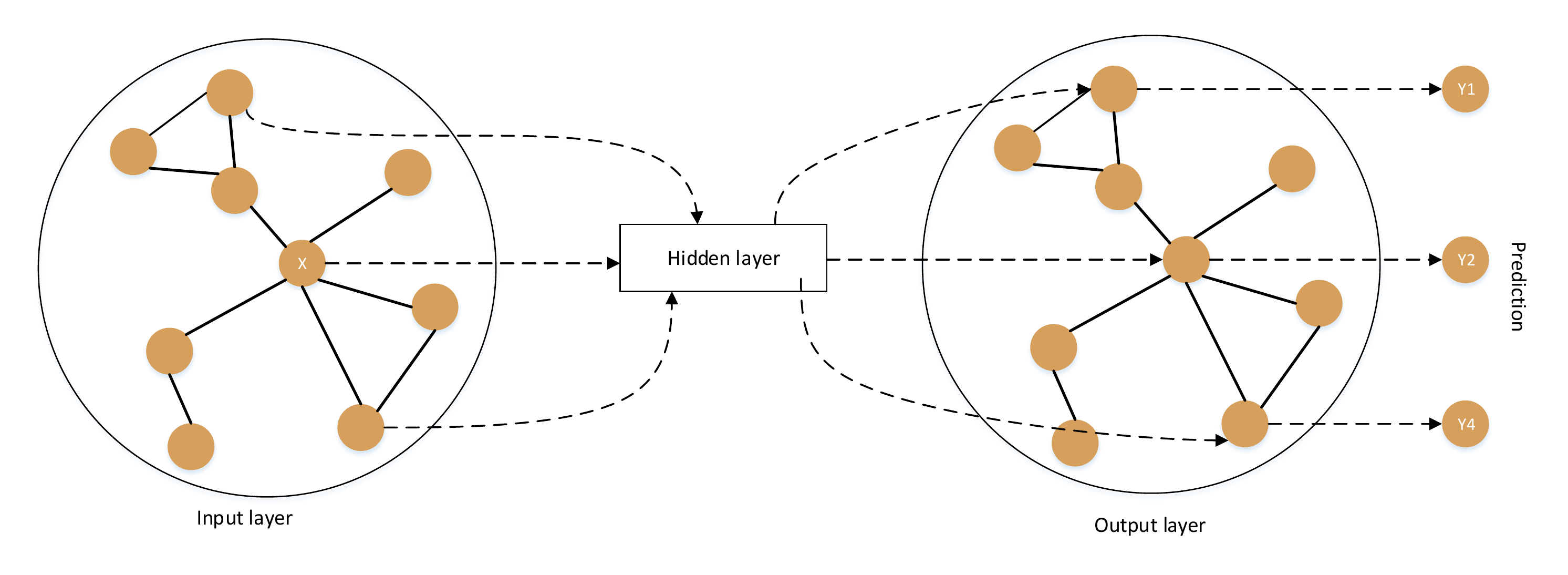}
	  \vspace{-1mm}
	  \caption{Schematic diagram of multi-layer graph convolutional network.}
	\label{fig_1}	  
	\vspace{-5mm}
\end{figure}

\section{Related Work}
\subsection{Graph Neural Networks}
A graph neural network (GNN) is a deep learning model for graph as a data structure, which has become a widely used graph representation learning method in recent years due to its excellent performance in several graph-based machine learning tasks and high interpretability ~\cite{zhu2020deep}. The concept of graph neural network was first introduced in 2005 by Gori et al.~\cite{gori2005new}, who designed a model for processing graph structure data by drawing on research results in the field of neural networks. In 2009, Scarselli et al.~\cite{scarselli2008graph} elaborated this model. Since then, new models and applications of graph neural networks have been proposed. In recent years, with the increasing interest in graph-structured data, the number of research papers on graph neural networks has shown a rapid increase, and the research directions and applications of graph neural networks have been greatly expanded. The paper~\cite{bronstein2017geometric} provides a review of deep learning methods in the field of graph-structured and streamlined data, focusing on placing the various methods described within a unified framework called geometric deep learning. The paper~\cite{zhang2020deep} classifies graph neural network methods into three categories: semi-supervised learning, unsupervised learning, and recent advances, and introduces, analyzes, and compares the methods according to their development history. The paper~\cite{zhou2020graph} introduces the original model, variants and general framework of graph neural networks, and classifies the applications of graph neural networks into structural, unstructured and other scenarios. The paper~\cite{wu2020comprehensive} proposes a new classification method for graph neural networks, focusing on graph convolutional networks, and summarizes the open source code and benchmarks of graph neural network methods for different learning tasks.

\subsection{Graph Neural Networks in Recommender Systems}
As an advanced deep learning-based graph representation learning method, GNN can capture the deep nonlinear features in the graph structure by using the neighbor structure of user and item nodes, i.e., historical interaction information, to generate effective feature representations for users and items~\cite{wei2021hierarchical}. In turn, it provides feature input for downstream recommendation tasks. However, in practice, it is found that the interaction data in most recommender systems are large in size and sparse, which poses a great challenge to the training efficiency of the model. To alleviate the data sparsity problem, RMG incorporates the interaction features contained in the review text and extracts the structural features of the interaction network using hierarchical attention networks~\cite{wu2020comprehensive}. In contrast, Multi-GCCF focuses more on the static features of users and items, constructs user relationship graphs and item relationship graphs using similarity, and applies graph convolutional neural networks as well as residual link learning in multiple graphs to obtain the final representation ~\cite{sun2019multi}. This class of models, by introducing or constructing additional data to provide higher quality feature inputs to GNNs, allows the quality of the model-generated representation to be improved, resulting in better recommendation results. Unlike the above methods, NGCF models simulate the process of information flow between users and items by constructing a forward propagation model based on message passing, which in turn mines to obtain deeper structural features, thus effectively improving the quality of the representation~\cite{wang2019neural}. In addition, from the perspective of model training efficiency, inspired by the GraphSAGE model, PinSage introduces a sampling strategy and adopts a distributed computing framework, which greatly improves the operational efficiency of GNN in large interaction networks.

\subsection{Multimodal recommendation}

In the multimodal recommendation , the main currently used recommendation models are divided into collaborative filtering-based video recommendation~\cite{liu2019using}, content-based video recommendation~\cite{yadalam2020career} and hybrid video recommendation~\cite{zheng2019unified}. YouTube~\cite{baluja2008video} in 2008 used User-Video-based graph tour algorithm is a kind of propagation diffusion of video labels on the graph of collaborative filtering algorithm, but only the video label information is considered, and there is a cold start problem~\cite{wei2021contrastive}. Mei et al.~\cite{mei2011contextual}  designed a contextual video recommendation system based on multimodal content relevance and user feedback, considering the different composition of the video and the different levels of user interest in different parts of the video, and seamlessly integrating multimodal relevance and user feedback through relevance feedback and attention fusion, but ignoring the role of user. Zhao et al.~\cite{zhao2013video} treated video recommendation as a ranking task, integrated users' personal background information and video information, and designed and implemented a user-adaptive multi-task ranking aggregation algorithm, but lacked the consideration of multimodal information relevance. In recent years, short videos have gradually become the mainstream trend in the social media era, and researchers have paid more and more attention to the study of short video recommendation scenarios. Users in the short video scenario have more unique characteristics, i.e., diverse and multi-level dynamic interests, which pose higher requirements for modeling user interests using multimodal information.Chen et al.~\cite{chen2018temporal} modeled users' historical behaviors to predict users' click-through rates for short videos, and proposed a temporal hierarchical attention mechanism on the category and item level (THACIL) network. Firstly, a time window is used to capture users' short-term interests, followed by a category-level attention mechanism to characterize users' different interests, as well as an item-level attention mechanism for fine-grained analysis of users' interests, and a forward multi-headed self-attention mechanism to capture long-term relevance. With the popularity of graph neural networks, Li et al. ~\cite{li2019routing} added users' multilevel interests to the user matrix, and then used graph-based sequential networks to model users' dynamic interests. And Wei et al.~\cite{MMGCN} designed the Multi-modalGraphConvolutionNetwork (MMGCN) framework. Based on the message passing idea of graph neural networks to construct multimodal representations of users and micro-videos, the higher-order connections between users and micro-videos in each modality are encoded using information propagation mechanisms to capture user preferences.Wang et al.~\cite{wang2019kgat} combine knowledge graphs and GAT graph attention networks to construct collaborative knowledge graphs that use attention mechanisms to learn user preferences and mine higher-order connections. The latest advancement in graph structure-based recommendation algorithms is the Multi-modalKnowledgeGraphAttentionNetwork (MKGAT), and Sun et al.~\cite{sun2020multi} further combined multi modal and knowledge graphs together, and proposed a multimodal graph attention technique to propagate information over a multimodal knowledge graph (MMKG) and use the obtained clustered embedding representation for recommendation to better enhance the recommender system.
\section{Methodology}
Given an original input feature matrix $X$ and an undirected graph $\mathcal{G}$ the GCN will perform the following interlayer propagation operations:
\begin{equation}
H^{(l+1)}=\sigma(D^{-\frac{1}{2}}AD^{-\frac{1}{2}}H^{(l)}W^{(l)}),
\end{equation}
where $A$ is the adjacency matrix of the graph $\Delta$ with self-loop. $D$ is the degree matrix of $A$, and $W^{(l)}$ is the trainable parameter of each layer. $\sigma(\cdot)$ represents the activation function, e.g.,$ReLU(\cdot)=max(0,\cdot)$. $H^{(l)}\in \mathbb{R}^{N\times D}$ is the hidden identity matrix of the layer.

The output of the last layer of the GCN is the final representation of the graph node $H^{(l)}$ in the original GCN work, which focus is on the semi-supervised classification problem of graph nodes. To accomplish this task, the feature matrix of the final output $H^{(l)}$  for each row of softmax operation. Let $Z=softmax(H^{(l)})\in \mathbb{R}^{N\times C}$ be the final output and $C$ indicates the number of final node categories. $Z$ is the matrix consisting of the type vectors predicted for each node. The weight parameters of the GCN are trained by minimizing the cross-entropy loss function of all labeled nodes:
\begin{equation}
L=-\sum_{l\in y}\sum^F_{f=1}Y_{lf}ln Z_{lf},
\end{equation}
where $y$ represents the set of nodes with labels. $Y_{lf}$ and $Z_{lf}$ denote the dimensions of the feature prediction values.

\subsection{Knowledge graph construction layer}

The user-microvideo bipartite diagram can be defined as:
\begin{equation}
\mathcal{G}_1=\{(u,y_{ui},i)|u\in U,y_{ui}\in Interaction,i\in I\},
\end{equation}
where $U$ is the user-set, $I$ is the micro-video-set, and $Interact$ is the set of user-micro-video interactions. If there are interactions between users and micro-videos, then there are connected edges between users and micro-videos, and vice versa.

The knowledge graph consisting of user and user-side auxiliary information can be defined as:
\begin{equation}
\mathcal{G}_2=\{(h,r,t)|h,r\in R_1,t\in \epsilon_1\},
\end{equation}
Where $\epsilon_1$ represents the set of entities composed of user and user auxiliary information, $R_1$ represents the set of relations between user and its auxiliary information, and $(h,r,t)$ represents the triad consisting of head entity, tail entity and relation between two entities, by which the knowledge graph is described.

The knowledge graph consisting of micro-video and micro-video-side multimodal auxiliary information can be defined as:
\begin{equation}
\mathcal{G}_3=\{(h,r,t)|h,r\in R_2,t\in \epsilon_2\},
\end{equation}
Where $\epsilon_2$ represents the set of entities composed of multinodal information, $R_2$ represents the set of relations between microvideo and its multimodal information, and $(h,r,t)$ represents the triad consisting of head entity, tail entity and relation between two entities, by which the knowledge graph is described.

Entity alignment between different knowledge graphs using micro-video-Entity for Alignment A and User-Entity for Alignment B. The formal definitions of A and B are as follows:
\begin{equation}
A=\{(i,e)|i\in I,e\in E\},
\end{equation}
\begin{equation}
B=\{(u,e)|u\in U,e\in E\},
\end{equation}
where I represents the set of microvideos. U represents the set of users, and E represents the set of entity.

After integrating the user-micro-video bipartite graph and the user-side and micro-video-side knowledge graphs, we can finally obtain the user-side collaborative knowledge graph $\mathcal{G}_u$ and the micro-video-side collaborative knowledge graph $\mathcal{G}_i$, where the user-side collaborative knowledge graph $\mathcal{G}_u$ takes the user node as the starting point of the relationship and points to the micro-video node with interaction, and then to the micro-video attributes, which are defined as:
\begin{equation}
\mathcal{G}_u=\{(h,r,t)|h,r\in R_u,t\in \epsilon_u\},
\end{equation}

n the user-side collaborative knowledge mapping, considering that there may be one or more interactions between users and micro-videos, suppose $r_6$ is a like relationship and $r_7$ is a favorite relationship, where users only like some micro-videos . The original KGAT algorithm does not distinguish fine-grained interaction relations, but when the subsequent coding of the entities in the knowledge graph is performed, it needs to calculate through the relationship space. If all fine-grained interaction relations are considered as the same, it is impossible to explore the deep interest behind the fine-grained interaction relations of users, so this paper adopts a composite relationship representation for the edges in the user-side collaborative knowledge graph to facilitate the subsequent Therefore, this paper adopts a composite relationship representation for the edges in the user-side collaborative knowledge graph to facilitate further processing in the subsequent knowledge graph coding layer. In the micro-video-side collaborative knowledge graph a, the micro-video node is used as the starting point of the relationship, which points to the user nodes with interaction relations and then to the user attributes, and is defined as:
\begin{equation}
\mathcal{G}_i=\{(h,r,t)|h,r\in R_i,t\in \epsilon_i\},
\end{equation}

\begin{figure}
	\centering
	  \includegraphics[width=0.5\textwidth]{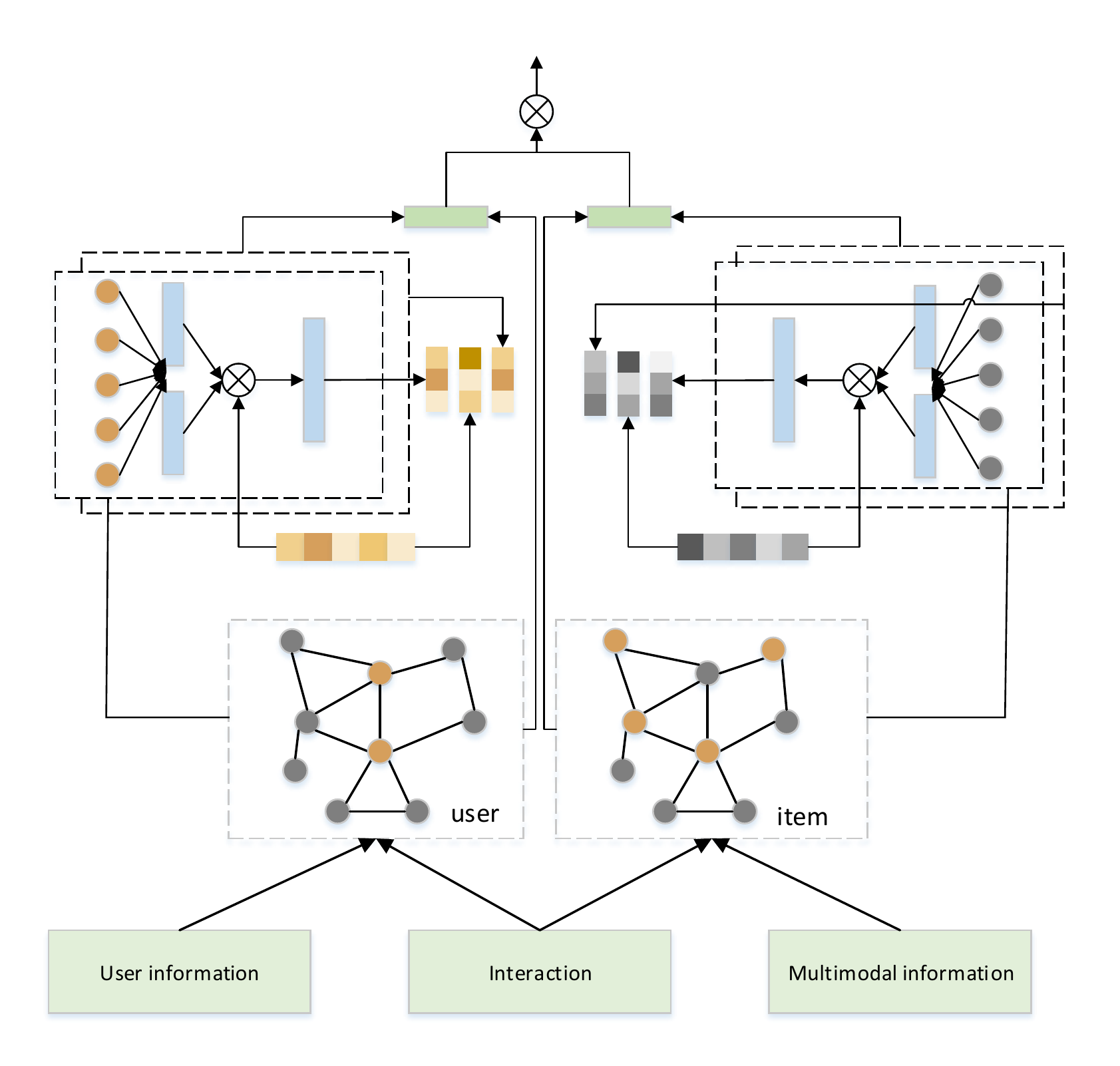}
	  \vspace{-1mm}
	  \caption{Schematic diagram of our model.}
	\label{fig_1}	  
	\vspace{-5mm}
\end{figure}

\subsection{Encoding layer}
This layer is mainly responsible for encoding the representation of the two knowledge graphs generated by the knowledge graph construction layer, i.e.$e_h,e_t \in \mathbb{R}^D, e_r\in \mathbb{R}^k$, embedding and modeling all entity vectors and relationship vectors in the knowledge graph. The core implementation principle of this layer is the TransR algorithm, where each entity and relationship in the knowledge graph is encoded and represented.Each entity and relationship in the knowledge graph is represented by encoding, where $e_h,e_t,e_r$ represent the encoding of head entity, tail entity and relationship respectively. TransR algorithm considers that the relationship spaces of different relationships should be different from each other, and therefore models them in the entity space and several different relationship spaces. For a relationship, the $e_h$ and $e_t$ entity vectors are projected in the corresponding relationship space such that$e_h^r+e_t=e_t^r$, where $e_h^r=W_r e^h$, where $W^r$ is a trainable transition matrix, and $e_h$ and $e_t$ are projected in a specific $r$ space using matrix multiplication to satisfy that the sum of the head entity vector and the relationship vector is as close to the tail entity vector as possible. The loss function of the knowledge graph encoding process is:
\begin{equation}
L_{Kg}=\sum_{(h,r,t,t')\in T} -ln\sigma(g(h,r,t')-g(h,r,t)),
\end{equation}
The TransR randomly selects other entities without r-connections as negative samples during the training process.

\subsection{Propagation layer}
The information dissemination phase learns the weights from neighboring nodes around a tail entity through an attention mechanism. The core formula is:
\begin{equation}
e_{N_h}=\sum_{(h,r,t\in N_h)}\pi(h,r,t)e_t,
\end{equation}
where $\pi(h,r,t)$ is the weight formula for measuring $(h,r,t)$. Its formula is defined as:
\begin{equation}
\pi '(h,r,t)=(W_r e_t)^T tanh(W_r e_h +e_t),
\end{equation}
\begin{equation}
\pi (h,r,t)=\frac{exp(\pi '(h,r,t))}{\sum_{(h,r',t'\in N_h)}exp(\pi '(h,r',t'))},
\end{equation}
Since the TransR algorithm exists the definition of head vector and relation vector summation to get the tail vector, if the summation operation on the r feature space can be satisfied, then the larger the weights obtained, the tanh in the formula can increase the nonlinear capacity. 

\subsection{Aggregation layer}
The information passed in the graph is aggregated and updated, i.e. the aggregated information of the neighbor nodes of the head entity and the vector of the head entity itself are aggregated and used as the new vector of the head entity itself.
\begin{equation}
f=LeakyReLU(W_1(e_h +e_{N_h}))+LeakyReLU(W_1(e_h \odot _{N_h})),
\end{equation}

Repeat the above steps for the current lth to obtain:
\begin{equation}
e_h^{(l)}=f(e_h^{(l-1)},e_{N_h} ^{(l-1)}),
\end{equation}
Stitching all the vectors generated in the l process into one vector.In this step, due to the existence of two knowledge graphs, two pairs of user vectors and micro-video vectors are obtained respectively, which are finally combined into one user vector and one micro-video vector by vector stitching operation.

\subsection{Prediction layer}
The predicted scores of user $u$ for micro-video $i$ are calculated by multiplying the synthesized two embedding vectors, and the TOP-K recommendation list is derived from the prediction results. The Pairwise BPR loss function used in the prediction stage is:
\begin{equation}
L_{CF}=\sum_{(u,i,j)\in O}-ln\sigma(y(u,i)-y(u,j)),
\end{equation}
where $O$ can define the overall loss function as the sum of the loss functions of each part. Considering that two types of knowledge graphs are used for common training in this paper, the overall loss function of the algorithm can be defined as:
\begin{equation}
L_{KGAT}=L_{KG_u}+L_{KG_i}+L_{CF}+\lambda||\Theta||^2_2,
\end{equation}
where $L_{KG_u}$ is the loss function of the user-side collaborative knowledge map at the knowledge map encoding layer, and similarly $L_{KG_i}$ is the loss function of the micro-video-side collaborative knowledge graph in this layer. Since the encoding vectors of the two knowledge graphs are spliced before the prediction stage, there is only one loss function $L_{CF}$ in the prediction process. Adding $\lambda||\Theta||^2_2$ can effectively avoid overfitting.

\section{EXPERIMENTS}
\subsection{Dataset}
In this paper, three datasets were selected, from which some data were chosen as test datasets, and the data size and multimodal information of each dataset are shown in Table 1.
\begin{itemize}
    \item MovieLens: This movie rating dataset has been widely used to evaluate collaborative filtering algorithms. While it is a dataset with explicit feedbacks, we follow the convention that transforms it into implicit data, where each entry is marked as 0 or 1 indicating whether the user has rated the item.
    \item Amazon Instant Video: The dataset consists of users, videos and ratings from Amazon.com. Similarly, we transformed it into implicit data and removed users with less than 5 interactions.
    \item Tiktok: It is published by Tiktok, a micro-video sharing platform that allows users to create and share micro-videos with duration of $3-15$ seconds. It consists of users, micro-videos and their interactions. The micro-video features in each modality are extracted and published without providing the raw data.
\end{itemize}
 
   \begin{table}
  \centering
  \caption{Dataset information.}
    %\vspace{0.1mm}
  \label{table_3}
  \setlength{\tabcolsep}{4.0mm}
  \begin{tabular}{|c|c|c|c|}
    \hline
    \textbf{Dataset}&\textbf{User}&\textbf{Item}&\textbf{Rating}\\
    \hline
    \textbf{MovieLens}&$45845$&$5986$&$839508$\\
    \hline
    \textbf{Amazon Video}&$37126$&$30648$&$583933$\\
    \hline
    \textbf{Tiktok}&$36656$&$76085$&$726065$\\
    \hline
    
  \end{tabular}
  \vspace{-2mm}
\end{table}

\subsection{Experiment setting}
The data set is divided into training set, testing set and validation set in the ratio of 8:1:1. Two widely used recommendation algorithm performance metrics, accuracy and recall were selected as the validation metrics for the algorithm comparison experiments in this paper. For the final implementation of the recommendation algorithm TOP-K recommendation, K = 10, the learning rate $\{0.1, 0.01, 0.001, 0.0001\}$ is set, the Gaussian matrix initialization parameters are used, and LeakyReLU is set as the activation function.

\subsection{Baselines}
NCF~\cite{he2017neural}: This model is based on a matrix decomposition framework,which utilizes an MLP to model the nonlinear relationship between users and products. The model achieves the best performance among hidden factor models due to the powerful representation capability of neural networks.
ACF~\cite{chen2017attentive}. This is the first framework that is designed to tackle the implicit feedback in multimedia recommendation. It introduces two attention modules to address the item-level and componentlevel implicit feedbacks. It introduces two attention modules to address the item-level and componentlevel implicit feedbacks.
NGCF~\cite{wang2019neural}: This modle represent a novel recommendation framework to integrate the user-item interactions into the embedding process. By exploiting the higher-order connectivity from user-item interactions, the modal encodes the collaborative filtering signal into the representation.

 \begin{figure*}
    \centering
    \subfigure[ ]{
      \includegraphics[width=0.4\textwidth]{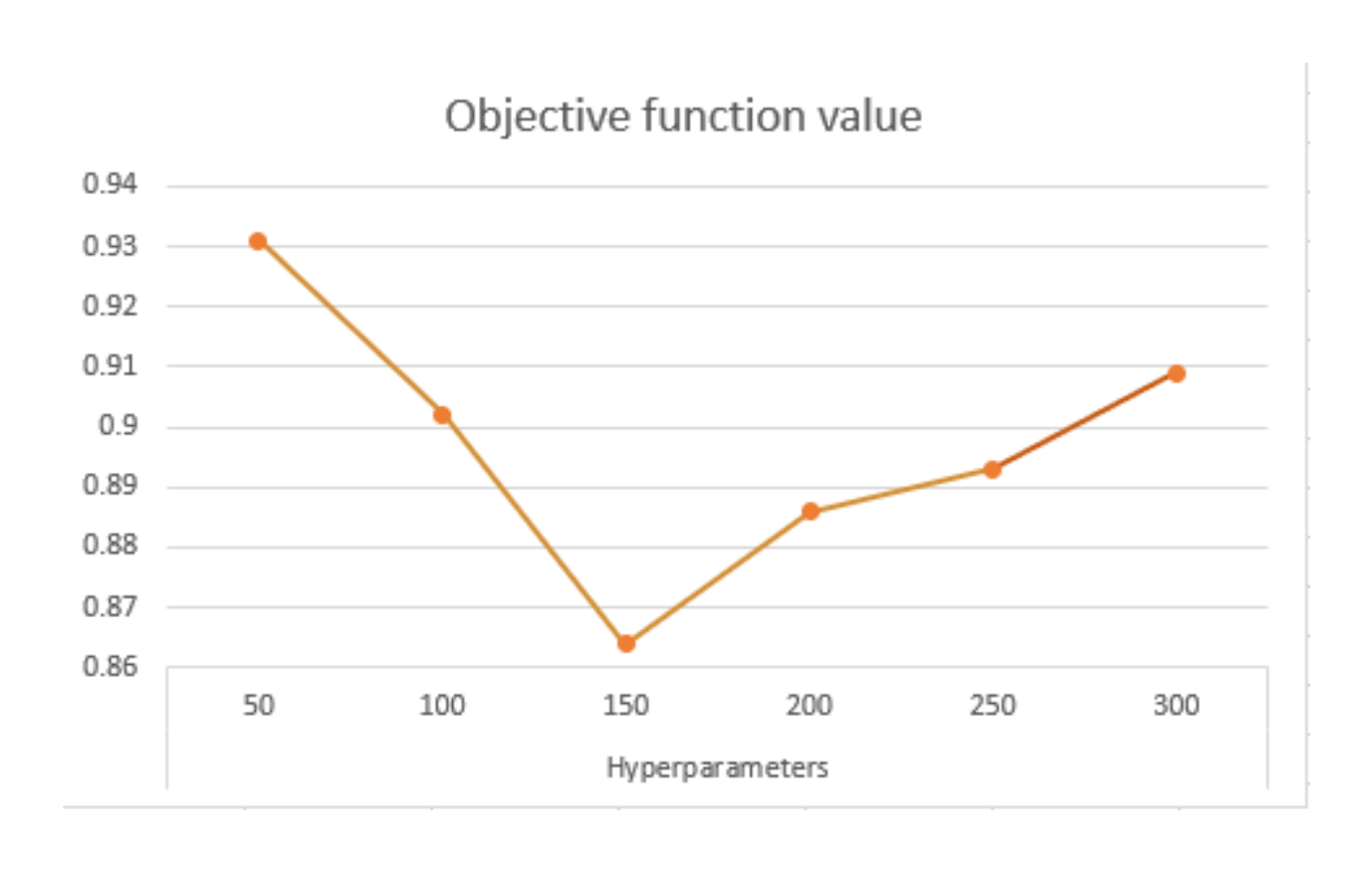}
      \label{fig_visualize_1_1}
    }
    \subfigure[ ]{
      \includegraphics[width=0.4\textwidth]{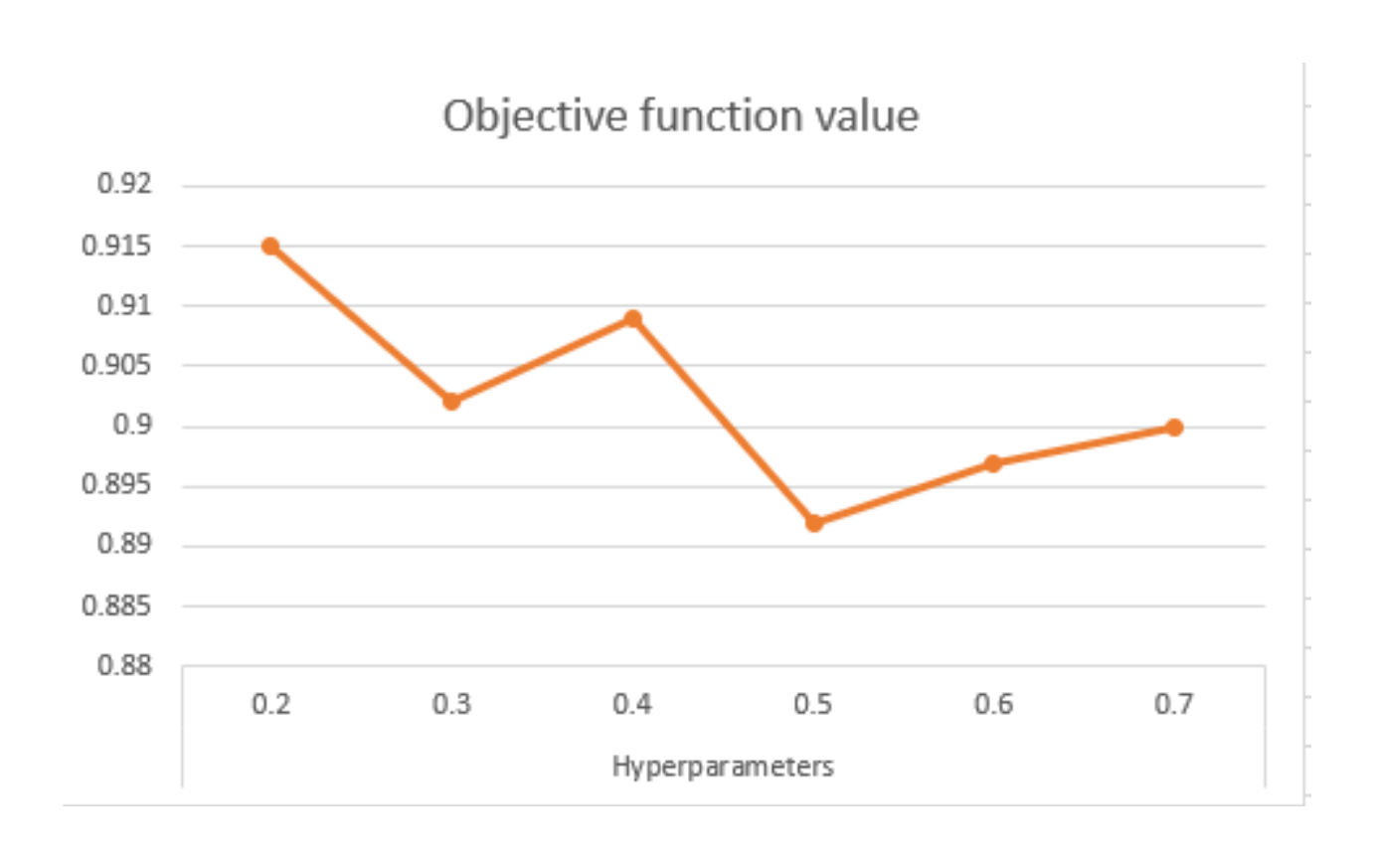}
      \label{fig_visualize_2_3}
    }

    \vspace{-10pt}
    \caption{Parameter adjustment result.}
    \label{fig_3}
    \vspace{-5pt}
 \end{figure*}

\subsection{Result Analysis}
On three datasets, our model proposed achieves better performance. Compared with other baselines, our model achieves higher recommendation accuracy and recall after introducing multimodal information to improve the network model. The model with the addition of the attention mechanism achieves improved results on most of the data sets, which can reflect that the addition of the attention mechanism does make the rating prediction more accurate. Since the original baseline model does not consider the importance of each dimension of user-item hidden features in the rating prediction, it also makes the model training stick to the task-oriented "graph representation learning", and the rating prediction is only used as a pseudo-task to evaluate the loss of the model to update the parameters of the graph representation learning model, which may also lead to the optimal user or item The optimal user or item representation cannot be achieved in the test set.
This paper also designed a comparison experiment on the effect of different network layers on the performance of the algorithm, and designed the number of network layers from 1 to 4 respectively. According to Tables 3-5, it can be seen that on the three data sets, the overall algorithm performance is best when the number of network layers is 3. The analysis shows that if the number of network layers gradually increases, i.e., more neighbor nodes are extended and errors are easily introduced, so the number of network layers When the number of network layers is greater than 3, the increase in the number of network layers reduces the recommendation accuracy instead.

   \begin{table}[htbp]
  \centering
  \caption{Performance comparison between our model and baselines.}
    %\vspace{0.1mm}
  \label{table_3}
  \setlength{\tabcolsep}{1.5mm}
  \begin{tabular}{c|cccccc}
    \toprule
    \multirow{2}*{Model}& \multicolumn{2}{c}{Movielens}&\multicolumn{2}{c}{Amazon Video}&\multicolumn{2}{c}{TikTok}\\
    % \hline
    % \cline{2-7}
     & Precision & Recall & Precision & Recall& Precision & Recall \\
    \hline
    ACF&0.1143&0.4631&0.1092&0.4783&0.1215&0.4176\\
    \hline
    NGCF&0.1238&0.4655&0.1107&0.4792&0.1221&0.4201\\
    \hline
    KGAT&0.1264&0.4678&0.1130&0.4861&0.1236&0.4233\\
    \hline
    Ours&0.1307&0.4721&0.1167&0.4957&0.1287&0.4308\\
    \hline
  \end{tabular}
  \vspace{-2mm}
\end{table}

The hyperparameters involved in the model are mainly the dimensionality k of the node state representation and the conditioning parameter in the scoring confidence-based neighbor sampling. Both of these hyperparameters affect the ability of the test model to converge to a local optimum. For example, the minimum value of the objective function at k = 50 is 0.931 and the maximum value at k = 150 is 0.915, which is a significant difference between the two "local optima". A similar situation can be seen in the other hyperparameter tests.  However, with settings in the range [0.2, 0.8], the objective function values do not vary too much and the maximum and maximum values do not vary too much. The difference between the maximum and minimum values is relatively small.
   \begin{table}[htbp]
  \centering
  \caption{Performance comparison for different network layers.}
    %\vspace{0.1mm}
  \label{table_3}
  \setlength{\tabcolsep}{1.5mm}
  \begin{tabular}{c|cccccc}
    \toprule
    \multirow{2}*{Model}& \multicolumn{2}{c}{Movielens}&\multicolumn{2}{c}{Amazon Video}&\multicolumn{2}{c}{TikTok}\\
    % \hline
    % \cline{2-7}
     & Precision & Recall & Precision & Recall& Precision & Recall \\
    \hline
    $L=1$&0.1223&0.4731&0.1129&0.4813&0.1226&0.4233\\
    \hline
    $L=2$&0.1257&0.4764&0.1170&0.4823&0.1242&0.4269\\
    \hline
    $L=3$&0.1272&0.4787&0.1191&0.4878&0.1287&0.4279\\
    \hline
    $L=3$&0.1264&0.4751&0.1185&0.4862&0.1256&0.4267\\
    \hline
  \end{tabular}
  \vspace{-2mm}
\end{table}
\section{Conclusion and Future Work}
In recent years, micro-videos have gradually become the mainstream trend in the social media era, and researchers have paid more and more attention to the study of micro-video recommendation scenarios. Users in micro-video scenarios have more unique characteristics, i.e., diverse and multi-level dynamic interests, which put forward higher requirements for modeling user interests using multimodal information. To investigate the problems in the micro-video recommendr system mentioned above, this paper designs a hybrid recommendation algorithm model based on multimodal information, introduces multimodal information and user-side auxiliary information in the network structure, fully explores the deep interest of users, measures the importance of each dimension of user and item feature representation in the scoring prediction task, makes the application of graph neural network in the recommendation system is improved by using an attention mechanism to fuse the multi-layer state output information, allowing the shallow structural features provided by the intermediate layer to better participate in the prediction task. The recommendation accuracy is improved compared with the traditional recommendation algorithm on different data sets, and the feasibility and effectiveness of our model is verified.

\bibliography{BIB/IEEEabrv, reference}

\begin{thebibliography}{10}
\providecommand{\url}[1]{#1}
\csname url@samestyle\endcsname
\providecommand{\newblock}{\relax}
\providecommand{\bibinfo}[2]{#2}
\providecommand{\BIBentrySTDinterwordspacing}{\spaceskip=0pt\relax}
\providecommand{\BIBentryALTinterwordstretchfactor}{4}
\providecommand{\BIBentryALTinterwordspacing}{\spaceskip=\fontdimen2\font plus
\BIBentryALTinterwordstretchfactor\fontdimen3\font minus
  \fontdimen4\font\relax}
\providecommand{\BIBforeignlanguage}[2]{{%
\expandafter\ifx\csname l@#1\endcsname\relax
\typeout{** WARNING: IEEEtran.bst: No hyphenation pattern has been}%
\typeout{** loaded for the language `#1'. Using the pattern for}%
\typeout{** the default language instead.}%
\else
\language=\csname l@#1\endcsname
\fi
#2}}
\providecommand{\BIBdecl}{\relax}
\BIBdecl

\bibitem{zhang2019deep}
S.~Zhang, L.~Yao, A.~Sun, and Y.~Tay, ``Deep learning based recommender system:
  A survey and new perspectives,'' \emph{ACM Computing Surveys (CSUR)},
  vol.~52, no.~1, pp. 1--38, 2019.

\bibitem{ahmed2017programmable}
I.~Ahmed, S.~Obermeier, S.~Sudhakaran, and V.~Roussev, ``Programmable logic
  controller forensics,'' \emph{IEEE Security \& Privacy}, vol.~15, no.~6, pp.
  18--24, 2017.

\bibitem{lecun2015deep}
Y.~LeCun, Y.~Bengio, and G.~Hinton, ``Deep learning,'' \emph{nature}, vol. 521,
  no. 7553, pp. 436--444, 2015.

\bibitem{zhu2020deep}
W.~Zhu, X.~Wang, and P.~Cui, ``Deep learning for learning graph
  representations,'' in \emph{Deep learning: concepts and architectures}.\hskip
  1em plus 0.5em minus 0.4em\relax Springer, 2020, pp. 169--210.

\bibitem{Yu}
X.~Yu, T.~Gan, Z.~Cheng, and L.~Nie, ``Personalized item recommendation for
  second-hand trading platform,'' in \emph{Proceedings of the 28th ACM
  International Conference on Multimedia}, 2020, pp. 3478--3486.

\bibitem{zhou2020graph}
J.~Zhou, G.~Cui, S.~Hu, Z.~Zhang, C.~Yang, Z.~Liu, L.~Wang, C.~Li, and M.~Sun,
  ``Graph neural networks: A review of methods and applications,'' \emph{AI
  Open}, vol.~1, pp. 57--81, 2020.

\bibitem{hamilton2017inductive}
W.~Hamilton, Z.~Ying, and J.~Leskovec, ``Inductive representation learning on
  large graphs,'' \emph{Advances in neural information processing systems},
  vol.~30, 2017.

\bibitem{chen2018fastgcn}
J.~Chen, T.~Ma, and C.~Xiao, ``Fastgcn: fast learning with graph convolutional
  networks via importance sampling,'' \emph{arXiv preprint arXiv:1801.10247},
  2018.

\bibitem{ying2018graph}
R.~Ying, R.~He, K.~Chen, P.~Eksombatchai, W.~L. Hamilton, and J.~Leskovec,
  ``Graph convolutional neural networks for web-scale recommender systems,'' in
  \emph{Proceedings of the 24th ACM SIGKDD international conference on
  knowledge discovery \& data mining}, 2018, pp. 974--983.

\bibitem{chen2017stochastic}
J.~Chen, J.~Zhu, and L.~Song, ``Stochastic training of graph convolutional
  networks with variance reduction,'' \emph{arXiv preprint arXiv:1710.10568},
  2017.

\bibitem{li2018deeper}
Q.~Li, Z.~Han, and X.-M. Wu, ``Deeper insights into graph convolutional
  networks for semi-supervised learning,'' in \emph{Thirty-Second AAAI
  conference on artificial intelligence}, 2018.

\bibitem{kipf2016variational}
T.~N. Kipf and M.~Welling, ``Variational graph auto-encoders,'' \emph{arXiv
  preprint arXiv:1611.07308}, 2016.

\bibitem{pan2018adversarially}
S.~Pan, R.~Hu, G.~Long, J.~Jiang, L.~Yao, and C.~Zhang, ``Adversarially
  regularized graph autoencoder for graph embedding,'' \emph{arXiv preprint
  arXiv:1802.04407}, 2018.

\bibitem{berg2017graph}
R.~v.~d. Berg, T.~N. Kipf, and M.~Welling, ``Graph convolutional matrix
  completion,'' \emph{arXiv preprint arXiv:1706.02263}, 2017.

\bibitem{wang2021dualgnn}
Q.~Wang, Y.~Wei, J.~Yin, J.~Wu, X.~Song, L.~Nie, and M.~Zhang, ``Dualgnn: Dual
  graph neural network for multimedia recommendation,'' \emph{IEEE Transactions
  on Multimedia}, 2021.

\bibitem{gori2005new}
M.~Gori, G.~Monfardini, and F.~Scarselli, ``A new model for learning in graph
  domains,'' in \emph{Proceedings. 2005 IEEE international joint conference on
  neural networks}, vol.~2, no. 2005, 2005, pp. 729--734.

\bibitem{scarselli2008graph}
F.~Scarselli, M.~Gori, A.~C. Tsoi, M.~Hagenbuchner, and G.~Monfardini, ``The
  graph neural network model,'' \emph{IEEE transactions on neural networks},
  vol.~20, no.~1, pp. 61--80, 2008.

\bibitem{bronstein2017geometric}
M.~M. Bronstein, J.~Bruna, Y.~LeCun, A.~Szlam, and P.~Vandergheynst,
  ``Geometric deep learning: going beyond euclidean data,'' \emph{IEEE Signal
  Processing Magazine}, vol.~34, no.~4, pp. 18--42, 2017.

\bibitem{zhang2020deep}
Z.~Zhang, P.~Cui, and W.~Zhu, ``Deep learning on graphs: A survey,'' \emph{IEEE
  Transactions on Knowledge and Data Engineering}, 2020.

\bibitem{wu2020comprehensive}
Z.~Wu, S.~Pan, F.~Chen, G.~Long, C.~Zhang, and S.~Y. Philip, ``A comprehensive
  survey on graph neural networks,'' \emph{IEEE transactions on neural networks
  and learning systems}, vol.~32, no.~1, pp. 4--24, 2020.

\bibitem{wei2021hierarchical}
Y.~Wei, X.~Wang, X.~He, L.~Nie, Y.~Rui, and T.-S. Chua, ``Hierarchical user
  intent graph network for multimedia recommendation,'' \emph{IEEE Transactions
  on Multimedia}, 2021.

\bibitem{sun2019multi}
J.~Sun, Y.~Zhang, C.~Ma, M.~Coates, H.~Guo, R.~Tang, and X.~He, ``Multi-graph
  convolution collaborative filtering,'' in \emph{2019 IEEE International
  Conference on Data Mining (ICDM)}.\hskip 1em plus 0.5em minus 0.4em\relax
  IEEE, 2019, pp. 1306--1311.

\bibitem{wang2019neural}
X.~Wang, X.~He, M.~Wang, F.~Feng, and T.-S. Chua, ``Neural graph collaborative
  filtering,'' in \emph{Proceedings of the 42nd international ACM SIGIR
  conference on Research and development in Information Retrieval}, 2019, pp.
  165--174.

\bibitem{liu2019using}
G.~Liu and X.~Wu, ``Using collaborative filtering algorithms combined with
  doc2vec for movie recommendation,'' in \emph{2019 IEEE 3rd Information
  Technology, Networking, Electronic and Automation Control Conference
  (ITNEC)}.\hskip 1em plus 0.5em minus 0.4em\relax IEEE, 2019, pp. 1461--1464.

\bibitem{yadalam2020career}
T.~V. Yadalam, V.~M. Gowda, V.~S. Kumar, D.~Girish, and M.~Namratha, ``Career
  recommendation systems using content based filtering,'' in \emph{2020 5th
  International Conference on Communication and Electronics Systems
  (ICCES)}.\hskip 1em plus 0.5em minus 0.4em\relax IEEE, 2020, pp. 660--665.

\bibitem{zheng2019unified}
D.~Zheng and J.~Huang, ``A unified probabilistic matrix factorization
  recommendation fusing dynamic tag,'' in \emph{2019 International Conference
  on Robots \& Intelligent System (ICRIS)}.\hskip 1em plus 0.5em minus
  0.4em\relax IEEE, 2019, pp. 69--72.

\bibitem{baluja2008video}
S.~Baluja, R.~Seth, D.~Sivakumar, Y.~Jing, J.~Yagnik, S.~Kumar,
  D.~Ravichandran, and M.~Aly, ``Video suggestion and discovery for youtube:
  taking random walks through the view graph,'' in \emph{Proceedings of the
  17th international conference on World Wide Web}, 2008, pp. 895--904.

\bibitem{wei2021contrastive}
Y.~Wei, X.~Wang, Q.~Li, L.~Nie, Y.~Li, X.~Li, and T.-S. Chua, ``Contrastive
  learning for cold-start recommendation,'' in \emph{Proceedings of the 29th
  ACM International Conference on Multimedia}, 2021, pp. 5382--5390.

\bibitem{mei2011contextual}
T.~Mei, B.~Yang, X.-S. Hua, and S.~Li, ``Contextual video recommendation by
  multimodal relevance and user feedback,'' \emph{ACM Transactions on
  Information Systems (TOIS)}, vol.~29, no.~2, pp. 1--24, 2011.

\bibitem{zhao2013video}
X.~Zhao, J.~Yuan, M.~Wang, G.~Li, R.~Hong, Z.~Li, and T.-S. Chua, ``Video
  recommendation over multiple information sources,'' \emph{Multimedia
  systems}, vol.~19, no.~1, pp. 3--15, 2013.

\bibitem{chen2018temporal}
X.~Chen, D.~Liu, Z.-J. Zha, W.~Zhou, Z.~Xiong, and Y.~Li, ``Temporal
  hierarchical attention at category-and item-level for micro-video
  click-through prediction,'' in \emph{Proceedings of the 26th ACM
  international conference on Multimedia}, 2018, pp. 1146--1153.

\bibitem{li2019routing}
Y.~Li, M.~Liu, J.~Yin, C.~Cui, X.-S. Xu, and L.~Nie, ``Routing micro-videos via
  a temporal graph-guided recommendation system,'' in \emph{Proceedings of the
  27th ACM International Conference on Multimedia}, 2019, pp. 1464--1472.

\bibitem{MMGCN}
Y.~Wei, X.~Wang, L.~Nie, X.~He, R.~Hong, and T.-S. Chua, ``Mmgcn: Multi-modal
  graph convolution network for personalized recommendation of micro-video,''
  in \emph{Proceedings of the 27th ACM International Conference on Multimedia},
  2019, pp. 1437--1445.

\bibitem{wang2019kgat}
X.~Wang, X.~He, Y.~Cao, M.~Liu, and T.-S. Chua, ``Kgat: Knowledge graph
  attention network for recommendation,'' in \emph{Proceedings of the 25th ACM
  SIGKDD international conference on knowledge discovery \& data mining}, 2019,
  pp. 950--958.

\bibitem{sun2020multi}
R.~Sun, X.~Cao, Y.~Zhao, J.~Wan, K.~Zhou, F.~Zhang, Z.~Wang, and K.~Zheng,
  ``Multi-modal knowledge graphs for recommender systems,'' in
  \emph{Proceedings of the 29th ACM International Conference on Information \&
  Knowledge Management}, 2020, pp. 1405--1414.

\bibitem{he2017neural}
X.~He, L.~Liao, H.~Zhang, L.~Nie, X.~Hu, and T.-S. Chua, ``Neural collaborative
  filtering,'' in \emph{Proceedings of the 26th international conference on
  world wide web}, 2017, pp. 173--182.

\bibitem{chen2017attentive}
J.~Chen, H.~Zhang, X.~He, L.~Nie, W.~Liu, and T.-S. Chua, ``Attentive
  collaborative filtering: Multimedia recommendation with item-and
  component-level attention,'' in \emph{Proceedings of the 40th International
  ACM SIGIR conference on Research and Development in Information Retrieval},
  2017, pp. 335--344.

\end{thebibliography}

\end{document}